# Magnetostructural Coupling at the Néel point in YNiO$_3$ Single Crystals

*Dariusz J. Gawryluk[1,†,\*], Y. Maximilian Klein[1,†,\*], Rebecca Scatena[2,3], Tian Shang[1,4], Romain Sibille[5], Denis Sheptyakov[5], Nicola Casati[6], Anthony Linden[7], Dmitry Chernyshov[8], Mirosław Kozłowski[9], Piotr Dłużewski[10], Marta D. Rossell[11], Piero Macchi[2,12] and Marisa Medarde[1,\*]*

[1]Laboratory for Multiscale Materials Experiments, Paul Scherrer Institut, 5232 Villigen PSI, Switzerland

[2]Department of Chemistry and Biochemistry, University of Bern, Freiestrasse 3, 3012 Bern, Switzerland

[3]Department of Physics, Clarendon Laboratory, University of Oxford, Parks Road, Oxford, OX1 3PU, United Kingdom

[4]Key Laboratory of Polar Materials and Devices (MOE), School of Physics and Electronic Science, East China Normal University, Shanghai 200241, China

[5]Laboratory for Neutron Scattering and Imaging, Paul Scherrer Institut, 5232 Villigen PSI, Switzerland

[6]Swiss Light Source, Paul Scherrer Institut, 5232 Villigen PSI, Switzerland

[7]Department of Chemistry, University of Zürich, Winterthurerstrasse 190, 8057 Zürich, Switzerland

[8]European Synchrotron Radiation Facility, CS40220, 38043 Grenoble CEDEX 9, France

[9]Łukasiewicz Research Network - Tele & Radio Research Institute, 11 Ratuszowa Street, 03-450 Warsaw, Poland

[10]Institute of Physics, Polish Academy of Science, Al. Lotników 32/46, 02-668, Warsaw, Poland

[11]Electron Microscopy Center, Empa, Swiss Federal Laboratories for Materials Science and Technology, 8600 Dübendorf, Switzerland

[12]Department of Chemistry, Materials and Chemical Engineering, Politecnico di Milano, Via Mancinelli 7, 20131 Milano, Italy

**ABSTRACT**

The recent discovery of superconductivity in infinite layer thin films and bulk Ruddlesden-Popper nickelates has stimulated the investigation of other predicted properties of these materials. Among them, the existence of magnetism-driven ferroelectricity in the parent compounds $R$NiO$_3$ ($R$ = 4$f$ lanthanide and Y) at the onset of the Néel order, $T_N$, has remained particularly elusive. Using diffraction techniques, we reveal here the existence of magnetostriction at $T_N$ in bulk YNiO$_3$ single crystals. Interestingly, the associated lattice anomalies are much more pronounced along the **b** crystal axis, which coincides with the electric polarization direction expected from symmetry arguments. This axis undergoes an abrupt contraction below $T_N$ that reaches $\Delta$**b**/**b** ~ - 0.01 %, a value comparable to those found in some magnetoresistive manganites and much larger than those reported for magnetism-driven multiferroics. This observation suggests a strong spin-lattice coupling in these materials, consistent with theoretical predictions. Using the symmetry-adapted distortion mode formalism we identify the main ionic displacements contributing to the lattice anomalies and discuss the most likely polar displacements below $T_N$. Furthermore, our data support symmetric superexchange as the most likely mechanism responsible for the magnetoelastic coupling. These results, that may be common to the full $R$NiO$_3$ family, provide new experimental evidence supporting the predicted existence of magnetism-driven ferroelectricity in $R$NiO$_3$ perovskites.

**INTRODUCTION**

Lattice modulations involving the self-organization of charges have attracted the interest of the scientific community since its existence was theoretically predicted by Wigner in 1934.[1] Known as charge density waves in metals,[2-6] these phenomena are usually designated by the generic name of charge order in semiconductors and insulators, where the electrons tend to localize in the proximity of the atomic positions or along the bonds.[7-10] Interestingly, if non-centrosymmetric magnetic order appears in the presence of a charge-ordered pattern, the coupling of both kinds of order may lead to polar charge displacements and give rise to ferroelectricity.[11] Charge-ordered $Ca_3CoMnO_6$, where ferroelectricity emerges at the onset of a space-symmetry breaking magnetic structure, was proposed to be an experimental realization of this mechanism.[12] However, clear experimental evidence supporting the predicted charge displacements at the origin of the polarization[13,14] has remained elusive to date.[15] Given the potential of this mechanism for achieving large polarization values close to room temperature (RT),[11] and the high current interest in materials with a strong coupling between magnetism and ferroelectricity for the development of low-power magnetoelectric applications,[16, 17] it is desirable to identify additional compounds where electric polarization could arise from the combination of charge and non-centrosymmetric magnetic orders.

Here we address this question by investigating the $YNiO_3$ perovskite, where space-symmetry breaking magnetic order develops below a Néel temperature $T_N \sim 153$ K in a pre-existing charge ordered pattern described by the centrosymmetric space group (SG) $P2_1/n$ .[18, 19] The charge order is characterized by a breathing distortion involving the existence of "large" and "small" $NiO_6$ octahedra (Fig. S1), and is associated with a metal-to-insulator transition (MIT) with $T_{MIT} \sim 580$ K ,[20, 21] (Fig. S2). As first reported in ref.[18], the unique, nominally $Ni^{3+}$ ($t_{2g}^6 e_g^1$) site present in the high-temperature metallic state (SG *Pbnm*) splits into two distinct Ni sites below $T_{MIT}$. These sites host expanded ("large") and contracted ("small") NiO$_6$ octahedra that alternate along the three pseudocubic perovskite axes, an observation that was interpreted as the signature of a partial $2Ni^{3+} \rightarrow Ni^{3+\delta} + Ni^{3-\delta}$ charge disproportionation accompanied by $Ni^{3+\delta}/Ni^{3-\delta}$ (Ni$^S$/Ni$^L$) charge order. At lower temperatures ($T < T_N$), the magnetic moments $\mu^L$ and $\mu^S$ associated to the two Ni sites order in a long-range pattern described by the magnetic propagation vector **k** = (1/2, 0, 1/2) (or **k** = (1/4, 1/4, 1/4) in the pseudocubic parent perovskite lattice), believed to be common to all *R*NiO$_3$ nickelates (*R* = 4f trivalent lantanide, Y$^{3+}$ and Tl$^{3+}$). This propagation vector, unique among perovskite oxides, involves a repetition of the Ni magnetic moment arrangement every four unit cells along the three pseudocubic crystal axes **a$_p$**. According to the literature, two magnetic structures (collinear ↑↑↓↓,[22] and non-colinear ↑→↓←,[23] with ↑ and ↑ representing the Ni magnetic moments $\mu^L$ and $\mu^S$ of the "large" and "small" NiO$_6$ octahedra), are compatible with the neutron powder diffraction data (Fig. S1). Interestingly, both are non-centrosymmetric and can potentially induce acentric displacements of charges (either ionic and/or electronic) below $T_N$ in order to minimize either the non-relativistic exchange energy $J_{ij}$ **S$_i$·S$_j$**, and/or the relativistic antisymmetric product $D_{ij}$ **S$_i$xS$_j$** of neighbouring spins **S$_i$** and **S$_j$**.[17,24] Although the details of the charge and magnetic orders are still subject of debate,[18, 22, 23, 25, 26 27-30] it is worth noting that group theory arguments, based exclusively on the symmetry of the paramagnetic SG *P2$_1$/n* and that of the magnetic propagation vector, predict a symmetry breaking from (non-polar) *P2$_1$/n* to (polar) *P2$_1$* below $T_N$ and the appearance of spontaneous electric polarization along the **b** crystallographic axis.[31]

The experimental verification of this prediction has been hampered to date by the difficult procedure required to synthesize high-quality bulk *R*NiO$_3$ compounds, which requires the use of elevated pressures (up to 60 kbar) and temperatures (up to 1000 °C).[20, 32-37] Very recently, we have successfully grown the first $R$NiO$_3$ single crystals with $R$ = Nd, Sm, Gd, Dy, Y, Ho, Er, and Lu using a novel route involving the use of molten salt fluxes and much lower oxygen gas pressure (2 kbar).[38] Such crystals were recently employed to carry out Raman spectroscopy measurements,[39] which revealed a symmetry decrease in YNiO$_3$ below $T_N$ consistent with the theoretical prediction. However, the nature of the associated structural modifications could not be established. Here we employ these crystals to investigate the evolution of the crystal structure across $T_N$ using diffraction techniques. Our analysis reveals the presence of pronounced lattice anomalies in the **b** crystal axis, consistent with the Raman results and with group theory arguments. It also uncovers the main ionic displacements contributing to the lattice anomalies, which point towards symmetric superexchange as the most probable mechanism at the origin of the magnetoelastic coupling. To conclude, we discuss the most likely polar displacements below $T_N$ and the possible origin of the new Raman modes observed at the onset of the Néel order.

**RESULTS AND DISCUSSION**

**Lattice response to the electronic and antiferromagnetic orders**

The YNiO$_3$ single crystals used in this work (Fig. 1A) were thoroughly characterized using a broad palette of techniques, as described in ref.[38] (see Materials and Methods section for a summary). Their metal-to-insulator and Néel temperatures, measured by heating using Differential Scanning Calorimetry ($T_{MIT}$) and SQUID magnetometry ($T_N$), were respectively 579.3 K and 152.7 K, in excellent agreement with the values previously reported for ceramic samples[18, 19] (Fig. S2). Because of their regular, prismatic habit and small size (between ~5 and ~30 μm, with some large exemplars reaching up to ~100 μm), they could be used as-prepared (i.e., uncrushed) for both single crystal and high-resolution powder diffraction measurements. Elemental maps obtained using energy-dispersive X-ray spectroscopy (EDX), shown in Fig. 1B, indicate a homogeneous chemical composition. Atomic-resolution high-angle annular dark-field scanning transmission electron microscopy (HAADF-STEM) images acquired along the [100], [010] and [310] directions of a representative YNiO$_3$ specimen, shown in Figs. 1C-H together with their Fourier transforms, reveal also a regular, defect-free cationic arrangement. This observation is supported by the resolution-limited rocking curve measured over the (004) reflection of a 50 × 50 × 75 μm crystal using neutron single crystal diffraction with λ = 1.178 Å (insert of Fig. 1A). Similar conclusions were obtained upon examination of other crystals, confirming their excellent overall quality.

Figs. 1I-K shows maps of the reciprocal space planes (*hk0*), (*h0l*) and (*0kl*) obtained at room temperature (RT) using laboratory X-ray single crystal diffraction (XRSCD) with λ = 1.5418 Å (Cu Kα). This technique is not ideally suited for detecting small lattice changes, such as the tiny deviation of the monoclinic angle from 90° reported in the literature.[18, 38] The *P2$_1$/n* symmetry was nevertheless confirmed by both high-resolution synchrotron X-ray (XRPD) and neutron powder diffraction (NPD) data, where the splitting of the reflections that are symmetry-equivalent in *Pbnm*, but not in *P2$_1$/n*, was clearly observed (Figs. 2B-C). The results of a joint (powder + single crystal) structural refinement conducted using this last SG are shown in Figs. 1L-O and summarized in Tables S1 and S2. The refined lattice parameters, atomic coordinates, anisotropic mean-square atomic displacements (reported here for the first time using the SG *P2$_1$/n*), and interatomic distances and angles are in very good agreement with those previously reported for policrystalline YNiO$_3$ samples,[40] further confirming the excellent match between the crystal structure and properties of our crystals, and the ceramic samples investigated in former studies.

Having demonstrated the excellent quality of our YNiO$_3$ single crystals, we used them to investigate the existence of a symmetry decrease below *T$_N$*. Strong experimental evidence in this sense was provided recently by Raman spectroscopic measurements,[39] which revealed the appearance of new modes upon entering the antiferromagnetic (AFM) phase. Since the total number of Raman-active modes allowed by the centrosymmetric SG *P2$_1$/n* of the insulating paramagnetic phase is 24, the observation of 30 YNiO$_3$ modes below $T_N$ clearly indicates that the low temperature SG should be different, and its symmetry lower. In order to check the predicted *P2$_1$/n* → *P2$_1$* symmetry-breaking sequence, we performed laboratory XRSCD measurements above and below $T_N$ with the aim of investigating the appearance of reflections forbidden in *P2$_1$/n* below this temperature. Unfortunately, their presence could not be established unambiguously due to the presence of *P2$_1$/n* systematic absence violations of the type (*h 0 l*) with (*h + l*) odd at all temperatures (Fig. S3). Given the high perfection of our crystals, we tentatively attribute these observations to multiple scattering (see discussion in the Materials and Methods section). The non-homogeneous temperature dependence of the forbidden reflection intensities suggests nevertheless that the existence of a lower symmetry space group below $T_N$ cannot be excluded from the present observations, but a definitive answer will need additional experimental work.

Complementary to the search for systematic absence violations in single crystals, we investigated the existence of lattice anomalies at the Néel temperature using high-resolution powder diffraction (neutron and synchrotron X-ray). In spite of the inherent Bragg reflection superposition, these techniques (in particular synchrotron XRPD) are extremely sensitive to small lattice variations. This is illustrated in Fig. 2, which shows the evolution of selected regions in the synchrotron XRPD patterns of YNiO$_3$ between 500 and 80 K, where small discontinuities are clearly visible in the positions of some Bragg reflections at $T_N$. These discontinuities are reflected in the temperature dependence of the lattice parameters, which display well-defined anomalies at $T_N$ (Fig. 3). In the case of *a*, the discontinuities are extremely weak at both $T_{MIT}$ and $T_N$. For *c*, which undergoes the sharpest change at $T_{MIT}$, the anomaly at $T_N$ is also very weak. For the monoclinic angle $\beta$ (> 90 degrees for T < $T_{MIT}$), a tiny discontinuity at $T_N$ is also visible. Conversely, a pronounced decrease is observed for the $b$ lattice parameter at the onset of the Néel order, with $\Delta b/b \sim 0.011$. These observations, never reported previously for YNiO$_3$, are consistent with the Raman findings.[39] Moreover, the enhanced structural response along the **b** crystal axis provides strong experimental support to the theoretical prediction by Pérez-Mato *et al.*,[31] which favours the SG *P2$_1$*, with a polar axis along **b**, as the most likely SG below $T_N$.

**Response of the mode amplitudes to the antiferromagnetic order**

We shall show now that the observed lattice anomalies at $T_N$ are connected with the occurrence of subtle ionic displacements below this temperature. In order to identify them, we describe here the distortion of the crystal structure as a superposition of symmetry-adapted distortion modes ("frozen phonons"), a formalism described in ref.[41] that has been employed in several recent studies for investigating structural phase transitions[42-45] (Materials and Methods). We use exclusively neutron diffraction data for the structural analysis due to the higher sensitivity of neutrons to light atoms (such as oxygen), negligible neutron absorption cross sections of the YNiO$_3$ constituents, and the larger temperature region investigated using this technique (10 K ≤ $T$ ≤ 600 K).

For the mode decomposition of the low temperature crystal structure (T < $T_N$), carried out with the *Amplimodes* software, [41] we use here the predicted polar SG *P2$_1$*. The thirty distortion modes allowed by this SG belong to sixteen different irreducible representations (*irreps*), reported here for the first time in Table S5 together with their basis vectors. The associated atomic displacements are illustrated in Fig. S4. The eight *irreps* labelled $R_4^+$, $M_3^+$, $X_5^+$, $R_5^+$, $M_2^+$, $R_1^+$, $M_5^+$ and $R_3^+$, with the capital letter indicating their **k**-vector in the first Brillouin zone of the cubic primitive $Pm\bar{3}m$ parent cell (Figs S4a and b), are common to the centrosymmetric paramagnetic group $P2_1/n$, active between $T_{MIT}$ and $T_N$. The amplitudes of their associated distortion modes (12 in total, Table S5) are thus expected to account for the main distortions of the crystal structure between these two temperatures. The remaining *irreps*, labelled $GM_4^-$, $GM_5^-$, $R_4^-$, $X_3^-$, $X_5^-$, $M_2^-$, $M_3^-$, and $M_5^-$, are of pure $P2_1$ symmetry, have a total of 18 associated distortion modes (Table S5), and are expected to be active only below $T_N$. Given the smaller size of the lattice anomalies at this temperature compared with those at $T_{MIT}$, the amplitudes of these modes are also expected to be modest.

We focus first on the temperature dependence of the mode amplitudes associated with the 8 *irreps* of $P2_1/n$ parentage, that we investigate using high-intensity NPD data (Materials and Methods). These data are not sufficiently sensitive to the existence of tiny deviations from the $P2_1/n$ symmetry below $T_N$. However, they allow an investigation of the evolution of the modes describing the main distortions of the structure. As shown in our previous study on PrNiO$_3$,[44] the appearance of new, $P2_1/n$-allowed modes below $T_{MIT} = T_N$ has an impact on *all* the pre-existing modes of $Pbnm$ symmetry, which undergo pronounced anomalies at the transition (Fig. 4). In YNiO$_3$, where a symmetry decrease from $P2_1/n$ to $P2_1$ below $T_N$ has been predicted, the presence of anomalies at $T_N$ in the pre-existing modes of $P2_1/n$ symmetry would be thus *a priori* expected.

The evolution of the mode amplitudes as a function of the reduced temperature $T^* = T/T_{MIT}$, refined using the SG $P2_1/n$ in the full insulating region ($T < T_{MIT}$), is shown in Figure 4 for YNiO$_3$. For comparison, the figure shows also the mode amplitudes of PrNiO$_3$, which were recently reported by some of us and refined using the same SG.[44] As expected from the growing distortion of the crystal lattice for the nickelates with small $R$-cations, the amplitudes of the main octahedral rotations $R_4^+$ and $M_3^+$ (around the **b** and **c** crystal axes, respectively) are much larger for YNiO$_3$. In contrast, the anomalies at $T_{MIT}$ are more pronounced for PrNiO$_3$, in line with the larger discontinuities in the lattice parameters reported for this nickelate.[44] The $R$-cation displacements of $X_5^+$ and $R_5^+$ symmetry are also significantly larger for YNiO$_3$, in particular along the **b** axis ($X_5^+$). This is also the case for the breathing mode $R_1^+$, whose amplitude at 10 K is ~ 0.16 Å, more than 16% larger than in the case of PrNiO$_3$ (~ 0.12 Å). We note also the abrupt, huge increase of the $R_1^+$ amplitude below $T_{MIT}$ for both nickelates (up to ~ 10 times larger than the anomalies observed in the other modes), which signals unambiguously that the breathing mode is the leading structural response to the electronic localization associated with the MIT.[44]

A further interesting observation concerns the different behavior of the $M_2^+$ mode in both nickelates, which involves an in-plane rhombic distortion of the NiO$_6$ octahedra (Figs. 4 and S4a). In the metallic state ($T > T_{MIT}$), the amplitude of this mode is very small for both PrNiO$_3$ and YNiO$_3$ (~ 0.01 Å). For PrNiO$_3$, its amplitude decreases upon entering the insulating state, reaching values close to 0 at 10 K.[44] For YNiO$_3$, in contrast, it rises suddenly below $T_{MIT}$ reaching ~ 0.06 Å at 10 K, which is the largest increase of a mode amplitude after that of the breathing mode (~ 0.16 Å). This unexpected observation reveals a different structural response of the early and late nickelates to the electronic localization. For large $R$-cations (such as Pr), the response consists of a moderate breathing distortion plus a substantial enhancement of the main octahedral rotations ($R_4^+$, $M_3^+$) and $R$-cation displacements ($X_5^+$, $R_5^+$). In contrast, for smaller $R$-cations (such as Y), the structure reacts with a larger breathing distortion, a much weaker enhancement of the octahedral rotations and $R$-cation displacements, and a rhombic deformation of the NiO$_6$ octahedra's basal plane ($M_2^+$) that is absent in the earlier nickelates. We note also that the amplitude of the Jahn-Teller distortion ($R_3^+$), forbidden in the metallic state, is very small (close to our experimental uncertainty for YNiO$_3$) and nearly constant for both nickelates below $T_{MIT}$.

After comparing the PrNiO$_3$ and YNiO$_3$ mode amplitudes, we focus now on the temperature dependence of the YNiO$_3$ modes of *P2$_1$/n* symmetry across $T_N$. The enlarged view of this region, shown in Figs. 4C-J, reveals that the only octahedral deformation showing a clear anomaly at the onset of the antiferromagnetic order is the Jahn-Teller distortion *R$_3^+$*, at least within our experimental uncertainty. In the case of the Y displacement along **a** (*R$_5^+$*), a tiny anomaly at $T_N$ breaks the quasi temperature-independent evolution of this mode amplitude. In contrast, the octahedral rotations around **b** (*R$_4^+$*) and **c** (*M$_3^+$*), and the Y displacement along **b** (*X$_5^+$*), have more pronounced temperature dependences, as well as clear anomalies at $T_N$, with amplitudes nearly constant (*X$_5^+$*) or even decreasing (*R$_4^+$*, *M$_3^+$*) below this temperature. This is reflected in the evolution of the average Ni-O-Ni superexchange angle, θ, and the Y bond-valence sums (Fig. 5), where small, but well-defined anomalies at $T_N$ are also observed. In the particular case of θ, the monotonic decrease of this angle observed upon cooling from 600 K is interrupted at $T_N$, and further reversed below this temperature. This promotes the stabilization of a structure with less pronounced octahedral rotations below $T_N$, in full agreement with the evolution of the R$_4^+$ and M$_3^+$ mode amplitudes.

The increase of the Ni-O-Ni superexchange angle below $T_N$ is a particularly interesting observation, because it encodes direct information about the mechanism at the origin of the magnetism-driven structural changes in YNiO$_3$. In general, these changes occur because the resulting magnetic interactions lead to a net energy decrease with respect to the undistorted configuration. In insulators, two kinds of magnetic interactions may drive the associated atomic shifts at temperatures as high as 150 K: the symmetric superexchange interaction $J_{ij}$ **S$_i$·S$_j$** between neighbouring spins **S$_i$** and **S$_j$**, and the relativistic antisymmetric Dzyaloshinskii-Moriya (DM) exchange interaction $D_{ij}$ **S$_i$xS$_j$**. The θ change across $T_N$ can affect both energy terms, and in the case of the perovskite structure, $\cos^2\theta$ and $\cos\theta$ dependences have been proposed for for $J_{ij}$ and $D_{ij}$, respectively.[46-48] To compare the impact of the observed $\theta$ changes on these terms, we approximate them as $J = <J_{ij}> = A\cos^2\theta$ and $D = <D_{ij}> = B\cos\theta$. We also assume that, as suggested by the large, negative Curie-Weiss temperature of YNiO$_3$ (-245 K),[38] both, $J$ and $D$, are AFM. For the $A$ and $B$ coefficients, unknown for YNiO$_3$, we employ the values reported in ref.[49] for the orthoferrite YFeO$_3$, whose crystal structure (SG *Pbnm*, the same as metallic YNiO$_3$) and average Fe-O-Fe angle ($\approx 145°$) are very similar to those of YNiO$_3$. According to this study, $J \approx$ 4.75 meV and $D \approx 0.12$ meV. This result, which reflects the much weaker, spin-orbit nature of the DM interaction, is nearly identical to those reported for other Fe-based perovskite oxides (LaFeO$_3$[49]; BiFeO$_3$[50]). Using the experimental $\theta$ angle and the abovementioned expressions, we obtain $A \approx 7.08$ meV and $B \approx 0.15$ meV. We can now employ these values and the experimental YNiO$_3$ $\theta$ angles above and below $T_N$ (147.2° and 147.4°, respectively) to calculate the changes in $J$ and $D$. We obtain $\Delta J \approx 80\ \Delta D$. Within the present approximations, the net energy decrease associated with the symmetric superexchange term across $T_N$ is thus almost two orders of magnitude larger than the one associated with the antisymmetric Dzyaloshinskii–Moriya term. Symmetric exchange has also been found to be more advantageous than the DM interaction in BiFeO$_3$, where the onset of the long-period spiral antiferromagnetic order at $T_N = 650$ K is also accompanied by subtle lattice anomalies and a change in the (pre-existing) polarization.[24] Interestingly, the discontinuities along the observed polarization direction in BiFeO$_3$ (the **c** axis) are much less pronounced than along the predicted polarization direction in YNiO$_3$ (the **b** axis). As shown in Fig. 3, this axis undergoes an abrupt contraction below $T_N$ that reaches $\Delta$**b**/**b** ~ - 0.01 %. This value is comparable to those observed in some magnetoresistive manganites ($\Delta l/l$ ~ - 0.01-0.05 %),[51,52] and much larger than those reported for magnetism-driven multiferroics, where the lattice anomalies at the paraelectric-ferroelectric transition are extremely weak or even absent.[53-56] This let us hypothesize that YNiO$_3$ may develop a sizable polarization, with a lower limit of the order of the reported change $\Delta P$ in BiFeO$_3$ below $T_N$ (up to 0.04 μC/cm$^2$).[24, 57, 58] The verification of this prediction, which might not be straightforward because of the small size of our crystals (≈ 40 × 40 × 100 μm at most), is outside the scope of the present investigation and will require specialized contacting tools, as well as additional experimental work.

We turn now to the distortion modes of pure $P2_1$ symmetry. We attempted to determine their amplitudes using high-resolution NPD data at 10 K with particularly high statistics (Materials and Methods). However, the reflection overlap inherent to the powder diffraction technique was still high because of the monoclinic symmetry of the unit cell, and we could not obtain any substantial improvement with respect to the high intensity data (the refined amplitudes of the $P2_1$ modes were of the order of the standard uncertainties). In the following we thus focus on the description of the new modes of pure SG $P2_1$ symmetry and, in particular, on those allowing the existence of polarization along the **b** axis. As shown in Table S5, the only distortion modes able to generate ferroelectricity along **b** are those labelled (13) to (17), which belong to the gamma-point *irreps* $GM_4^-$ and $GM_5^-$. Their ionic displacements are schematically shown in Fig. 6. We note that modes (13) and (14) from $GM_4^-$ involve exclusively "acoustic-like" displacements of the Y and Ni cations along the **b** axis. In contrast, the oxygen distortion modes (15) and (16) from $GM_4^-$ and (17) from $GM_5^-$ have more complex displacement patterns involving both "acoustic" and "optic-like" displacements. We note also that none of them coincide with the alternation of long and short distances between "large" Ni and "small" Ni planes predicted by some authors to occur along the [111] direction of the pseudocubic perovskite lattice.[11]

The derivation of the full set of symmetry-allowed distortion modes of the $P2_1$ SG, reported here for the first time, may also be of interest for the identification of the new Raman modes observed below $T_N$, which could not be assigned in ref.[39]. Such modes, *a priori* optic in nature and covering energies between ~ 5 and ~ 70 meV, suggest the participation of both light (oxygen) and heavier cations (yttrium and nickel) in the new atomic motions below $T_N$. In Table S5, we identify sixteen optic modes of pure $P2_1$ symmetry. Seven of them involve heavy atoms: (19), (20), (22), (26) and (29) nickel; (25) and (28) yttrium. If the low temperature SG is $P2_1$, these modes could be related to the new modes with energies below 30 meV. In contrast, the modes labelled (15), (16), (17), (18), (21), (23), (24), (27) and (30) involving exclusively oxygen atoms could be related to the new modes observed at higher energies. The only purely acoustic modes are (13) and (14), which involve, respectively, displacements of Y and Ni along **b**, together with some components of the oxygen modes (15), (16) and (17). These modes are most probably not observed in the reported Raman spectra because of their low expected energies. However, they could provide an alternative origin of the lowest energy mode reported in ref. [39], which was tentatively assigned to a bimagnon bound state. A full identification of the new Raman modes observed below $T_N$ is nevertheless out of the scope of this study and will require additional experimental and theoretical work.

**CONCLUSIONS**

The main outcome of the present investigation is the experimental observation of a pronounced lattice response to the non-centrosymmetric antiferromagnetic order in $YNiO_3$. Such a response, theoretically postulated[11, 31] and supported by recent Raman measurements on this material,[59, 39] was not substantiated experimentally to date. Using a combination of diffraction techniques, we reveal the presence of lattice anomalies that are particularly pronounced along the **b** crystal axis of $YNiO_3$. This observation is in excellent agreement with group theory arguments, which predicted a decrease of the crystal symmetry from non-polar $P2_1/n$ to polar $P2_1$ and the appearance of electric polarization along this crystallographic direction. Using the symmetry-adapted distortion mode formalism for the analysis of the neutron powder diffraction data, we show that the lattice anomalies are connected with the occurrence of subtle ionic displacements involving (at least) the Y and the O sublattices. The observation of a clear increase in the Ni-O-Ni superexchange angle below $T_N$ also provides direct information about the mechanism responsible for the magnetoelastic coupling, suggesting that the low-temperature crystal structure promotes a magnetic configuration more favorable from the point of view of the symmetric superexchange than from that of the antisymmetric Dzyaloshinskii-Moriya interaction. Although our diffraction data could not confirm the predicted $P2_1/n \rightarrow P2_1$ symmetry breaking below $T_N$, we employed the symmetry-adapted distortion mode formalism to determine the polar displacements allowed by SG $P2_1$. Then, we used them to discuss the possible origin of the new Raman modes observed at the onset of the Néel order. These results, which could be common to the other members of the RENiO$_3$ family,[39, 60] provide strong experimental support for the predicted existence of ferroelectricity in rare-earth Ni perovskites. They should also contribute to an improvement in the understanding of the complex physics of these highly-correlated oxides and promote further experimental work aimed at establishing the ground state symmetry of these materials and the presence/absence of spontaneous polarization below $T_N$.

**MATERIALS AND METHODS**

**Crystal growth at 2000 bar oxygen pressure and primary characterization**

The growth and physicochemical characterization of the YNiO$_3$ single crystals used in this work has been reported previously[38]. We employed an innovative method based on the use of highly reactive eutectic salt mixtures as fluxes, a temperature gradient, and 2000 bar oxygen gas pressure, delivered by a non-commercial apparatus[61] at the Paul Scherrer Institute (PSI). The crystal habit, size distribution, and elemental composition were characterized using field-emission scanning electron microscopy (FE-SEM) coupled with energy-dispersive X-ray spectroscopy (EDX). The measurements were performed on a JEOL JSM-7600F microscope located at the Tele & Radio Research Institute in Warsaw, Poland. The device operated with a 15 keV incident electron beam energy, and was equipped with a SE (Secondary Electron) detector. For scanning transmission electron microscopy (STEM) investigations, carried out at the Swiss Federal Laboratories for Materials Science and Technology (Empa), the samples were prepared by focused-ion-beam milling with a FEI Helios NanoLab 660 G3 UC instrument. Then, high-angle annular dark-field (HAADF) STEM images were acquired with a FEI Titan Themis microscope equipped with a spherical-aberration probe corrector (CEOS DCOR) at 300 kV. A probe semi-convergence angle of 25 mrad was set in combination with an annular semi-detection range of the HAADF detector of 72–200 mrad. The phase purity and RT crystal structure were initially investigated using laboratory powder and single crystal X-ray diffraction. Differential scanning calorimetry and SQUID magnetometry were employed, respectively, to determine the metal−insulator and antiferromagnetic order temperatures.

**Synchrotron X-ray and neutron powder diffraction**

Synchrotron X-ray powder diffraction measurements (XRPD) were performed at the Swiss Light Source (SLS) of the PSI in Villigen, Switzerland. The $YNiO_3$ crystals were loaded in quartz capillaries ($D$ = 0.1 mm) and measured in transmission mode with a rotational speed of ~2 Hz at the Materials Science Beamline (MSBL) using a wavelength of $\lambda$ = 0.7750 Å.[62] The primary beam was vertically focused and slitted to about 300 × 4000 $\mu m^2$. Several powder diffraction patterns were recorded at temperatures between 80 and 500K using an Oxford Instruments nitrogen Cryojet cooler and a Mythen II 1D multistrip detector with energy discrimination ($2\theta_{max} = 120°$, $2\theta_{step} = 0.0036°$).[63] The measurements were performed by heating to the next desired temperature, waiting for about three minutes for temperature stabilization, and then acquiring the data, typically over 15 minutes. The wavelength, zero offset and instrumental resolution function were determined using Si (NIST SRM 640d) and LaB$_6$ (NIST SRM 660c) reference powder samples. All data were corrected for absorption effects by including calculated $\mu R$ values in the Rietveld refinements.

Neutron powder diffraction measurements (NPD) were carried out at the Swiss Neutron Source SINQ of the PSI in Villigen, Switzerland. The sample (about 2g of YNiO$_3$ microcrystals from the same batch) was placed inside a cylindrical vanadium can ($D = 0.6$ cm, $H = 5$ cm). Several neutron diffraction patterns were recorded at temperatures between 10 K and 600 K on the powder diffractometer HRPT [64] (Ge (533), $2\theta_{max} = 160°$, $2\theta_{step} = 0.05°$, $\lambda = 1.494$ Å, high-intensity mode with primary collimation $\alpha_1 = 24$') using successively a helium cryostat (10 K $\leq T \leq$ 300 K) and a furnace (300 K $\leq T \leq$ 600 K). The measurements were performed by heating to the next desired temperature, waiting for about five minutes for temperature stabilization, and then acquiring the data, typically over 3 - 4h. An additional pattern using the high-resolution mode ($\alpha_1 = 6$') and improved statistics (~ 24 h) was recorded at 10 K. The background from the sample environment was minimized using an oscillating radial collimator. The exact wavelength was refined from combined fits of the HRPT and MSBL powder diffraction data at RT. All data were corrected for absorption effects by including experimentally determined $\mu R$ values in the Rietveld refinements.

**Laboratory and synchrotron X-ray single crystal diffraction**

YNiO$_3$ single crystals with sizes between 20 and 100 μm (longest edge) were mounted in either glass capillaries ($D = 30$ μm) or on Kapton MicroMounts (MiTeGen LLC). Laboratory single crystal X-ray diffraction (SCXRD) data acquisitions at RT were conducted at the Department of Chemistry of the University of Zurich, Switzerland, using a Rigaku Oxford Diffraction SuperNova Diffractometer.[65] The device was equipped with a CCD area-detector, and two micro-focus X-ray sources delivering a choice of Mo K$\alpha$ ($\lambda$ = 0.7107 Å) or Cu K$\alpha$ ($\lambda$ = 1.5418 Å) radiation.

Further data acquisitions were made at the Department of Chemistry and Biochemistry of the University of Bern, Switzerland and at the Physics Department of the University of Oxford, United Kingdom. Measurements as a function of temperature and of X-ray wavelength were carried out with different Rigaku Oxford Diffraction Diffractometers, all equipped with Oxford Cryosystems Cryostream flow cryostats. Cu K$\alpha$ ($\lambda$ = 1.5418 Å) and Mo K$\alpha$ ($\lambda$ = 0.7107 Å) SuperNova instruments were equipped with CCD area detectors, whilst an Ag K$\alpha$ ($\lambda$ = 0.5608 Å) data collection at 110 K was performed with a Synergy-S diffractometer equipped with a 2D single photon counting hybrid pixel detector.

The laboratory data sets were complemented with measurements at the BM01 end station of the Swiss-Norwegian Beam Lines (SNBL) of the European Synchrotron Radiation Facility (ESRF), Grenoble, France, using a Pilatus2M detector (Dectris),[66] an Oxford Cryosystems Cryostream 700+ nitrogen cooler, and a wavelength of $\lambda$ = 0.53842 Å. Data acquisitions for crystal structure determination were made at several temperatures above and below $T_N$ = 152.6 K.

The data reduction was performed with the *CrysAlisPro*[65] package for all datasets. The intensities were corrected for Lorentz and polarization effects, and an empirical absorption correction using spherical harmonics[65] was applied. For the larger single crystals measured at the University of Bern and the University of Oxford, analytical absorption corrections were applied. Equivalent reflections were merged according to the different space groups and used for crystal structure determination and refinement, as well as for combined (powder plus single crystal data) structure refinements.

**Single crystal neutron diffraction**

Single-crystal neutron diffraction data were collected using the Zebra instrument at the Swiss Spallation Neutron Source SINQ in Villigen PSI, Switzerland. A YNiO$_3$ single crystal with size approximately 0.05 × 0.05 × 0.075 μm was mounted on a conical Al pin, and centred with translation stages inside a Eulerian cradle equipped with a helium closed-cycle refrigerator. The beam of thermal neutrons was monochromated using the (311) reflection of germanium crystals, yielding a neutron wavelength of 1.178 Å. Data were measured using a single $^3$He-tube detector in front of the sample whose slits were appropriately adjusted. The incoming neutron beam was slitted down to 0.4 mm horizontal and 16 mm vertical (vertically focused beam), approximately 300 mm before the sample position. A few nuclear Bragg reflections were collected using ω-scans. A representative rocking curve for the (004) nuclear Bragg reflection of YNiO$_3$ is presented in the insert of Figure 1A. The curve is resolution limited.

**Combined (single crystal + powder diffraction) data analysis**

The atomic coordinates and anisotropic mean-square displacements at RT were determined from joint refinements of the powder (HRPT and MSBL) and single crystal (Mo Kα and Cu Kα, University of Zurich) diffraction data, as implemented in the *FullProf Suite* package.[67] The obtained values are summarized in Table S1. Selected interatomic distances and bond angles are shown in Table S2 together with the Y, Ni and O Bond-Valence Sums at RT, as defined by Brown and Altermatt.[68, 69]

**Powder diffraction data analysis**

The evolution of the lattice parameters between 80 and 500 K was first determined from the analysis of the MSBL data. The evolution of the crystal structure between 10 and 600 K was subsequently investigated using neutron diffraction data and the symmetry-adapted distortion mode formalism,[41] as implemented in the *FullProf Suite* package.[67] To be consistent with our previous work and with most theory studies, we employed the same unit cell origin and the same *irrep* labelling as used in ref. [44] for the SG of the insulating paramagnetic phase ($P2_1/n$, $T_{MIT} > T > T_N$). Note that this choice differs from the convention used in most experimental papers, where Ni is at (1/2 0 0). The lattice transformation from the high-symmetry ($Pm\bar{3}m$) into the low-symmetry setting ($P2_1/n$) is $\mathbf{a} = \mathbf{a_p} - \mathbf{b_p}, \mathbf{b} = \mathbf{a_p} + \mathbf{b_p}, \mathbf{c} = 2\mathbf{c_p}$; (0, 0, 0), where the last vector indicates the origin of the low-symmetry cell (**a, b, c**) containing four formula units ($Z = 4$) expressed using the basis of the high-symmetry perovskite cell ($\mathbf{a_p}, \mathbf{b_p}, \mathbf{c_p}$) with $Z = 1$. For the mode decomposition in the insulating antiferromagnetic phase (T < $T_N$), we used a metrically equivalent cell with $Z = 4$ and the SG $P2_1$. Since the origin is undefined because of the lack of a symmetry centre in this SG, we chose it in such a way that the unit cell is compatible with the $P2_1$ standard setting in the International Tables of Crystallography, Volume A.[70] The lattice transformation from ($Pm\bar{3}m$) to ($P2_1$) is $\mathbf{a} = \mathbf{a_p} - \mathbf{b_p}, \mathbf{b} = \mathbf{a_p} + \mathbf{b_p}, \mathbf{c} = 2\mathbf{c_p}$; (-1/4, -3/4, 1/2), see also Table S3. As usual in the structural refinements of non-centrosymmetric space groups, one mode amplitude was fixed as a reference. Here, we fixed the amplitude of the mode *(13)*, belonging to *irrep GM4⁻*, that involves in-phase Y displacements along the **b** crystal axis (Table S5).

**Single crystal diffraction data analysis**

In contrast to the powder diffraction results, the lattice metrics remain compatible with an orthorhombic space group for all data sets. Yet single crystal X-ray diffraction is not an ideal technique for detecting small changes in the lattice parameters. This is due to the combined effect of divergent diffracted X-rays and crystal mosaicity, which results in large diffraction spots. Moreover, in the particular case of a *Pbnm* → *P2₁/n* transition, the reduction from the *mmm* to *2/m* Laue class implies inherently an intergrowth twinning about the crystal plane (0 0 1) or (1 0 0). The few indications of a symmetry reduction to monoclinic come from the violations of systematic absence rules for the *b* glide plane of *Pbnm*. However, it should be considered also that the *n* glide plane is associated with a number of systematic absence violations, albeit less. In both cases, the systematic absence violations are more evident for large crystals than for smaller ones, and they are not visible in the powder diffraction data. This reinforces the idea that systematic absence violations are due to the breakdown of the kinematic approximation, as expected for crystals of a high perfection. The agreement among symmetry equivalent reflections is slightly better if we assume a monoclinic *2/m* Laue class with the twofold axis parallel to **b**, rather than orthorhombic or any other monoclinic subclass (with a twofold axis parallel to **a** or **c**). However, this would not be a sufficient argument for choosing a monoclinic space group. Note that structural models refined in space group *P2₁/n* (with mirror plane twinning) and in *Pbnm* are almost equivalent in terms of agreement statistics between observed and calculated structure factors. Thus, the main proof of the monoclinic nature of the lattice comes from powder diffraction measurements (Figs. 1 and 2), where a clear splitting is observed between peaks which are symmetry-equivalent in the orthorhombic *mmm* Laue class, but not in *2/m*.

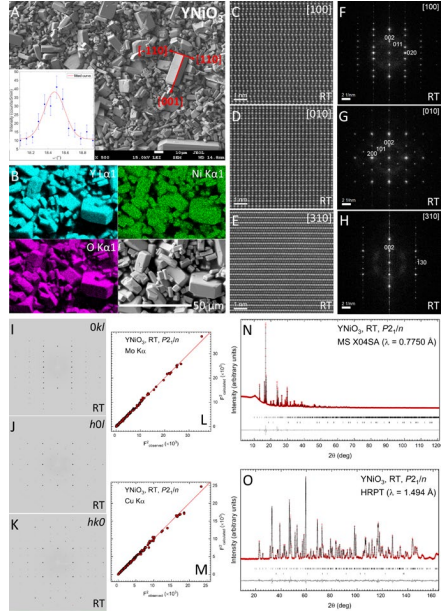

**Figure 1.** (A) Scanning Electron Microscopy image of YNiO$_3$ single crystals showing their habit and size distribution. Insert: Rocking curve over the (004) reflection measured at RT on a 50 × 50 × 75 μm specimen using neutron single crystal diffraction (λ = 1.178 Å). (B) Elemental maps (Y, Ni and O) obtained by means of EDX together with the SEM image of the mapped YNiO$_3$ single crystals. (C-E) Atomic-resolution HAADF-STEM images acquired at RT along the [100], [010] and [310] directions together with their corresponding Fourier transforms (F-H). (I-K) Reciprocal (*0kl*), (*h0l*), and (*hk0*), lattice planes reconstructed from laboratory XRSCD (Cu Kα) at RT for a YNiO$_3$ single crystal. (L-M). The agreement between the observed and calculated squared structure factors obtained from RT Mo Kα and Cu Kα datasets are shown, respectively, in panels (L) and (M). (N-O) Agreement between the observed and calculated powder diffraction data at RT, obtained from combined structural refinements (NPD, XRPD plus XRSCD with Mo Kα and Cu Kα) using the *FullProf Suite* software. In the Rietveld fits of panels (N) and (O) the red crosses and black lines correspond, respectively, to the experimental and calculated patterns, the first row of vertical ticks indicates the positions of the Bragg reflections allowed by the SG *P2$_1$/n*, while the second row represents those of a minor (< 1% in weight) NiO impurity

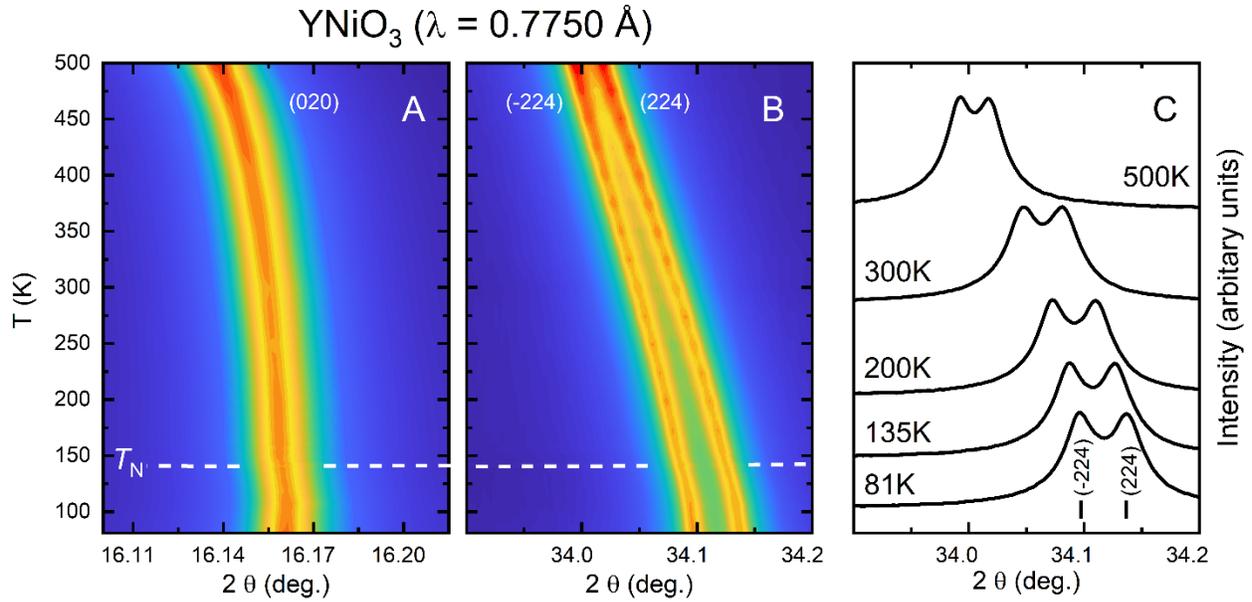

**Figure 2.** (A, B) Contour maps showing the temperature dependence of the position and the intensities of selected Bragg reflections for $YNiO_3$. The measurements were performed by heating at the X-ray powder diffractometer of the Materials Science X04SA Beamline (SLS, Switzerland). The white dashed lines highlight the lattice anomalies at the antiferromagnetic ($T_N$) transition temperature. (C) Monoclinic splitting of a selected reflection pair at different temperatures.

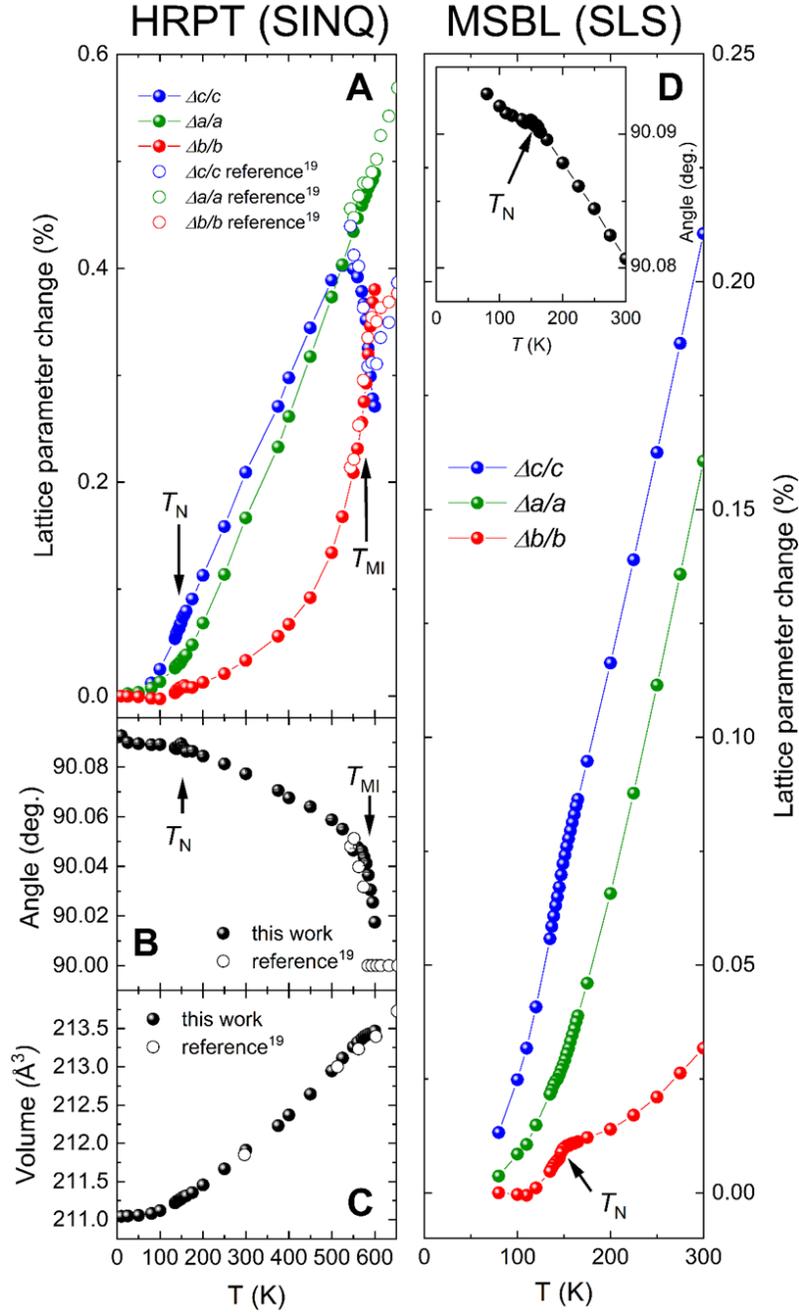

**Figure 3.** (A) Relative change (%) of the *a, b* and *c* lattice parameters as a function of temperature with respect to their values at 10 K, as determined from NPD data. Full symbols: this work. Open symbols: ref.[19] (B, C). Thermal evolution of the monoclinic angle $\beta$ and the unit cell volume *V* (from NPD). (D) Detailed view of the region around $T_N$, as obtained from the analysis of synchrotron XRPD data.

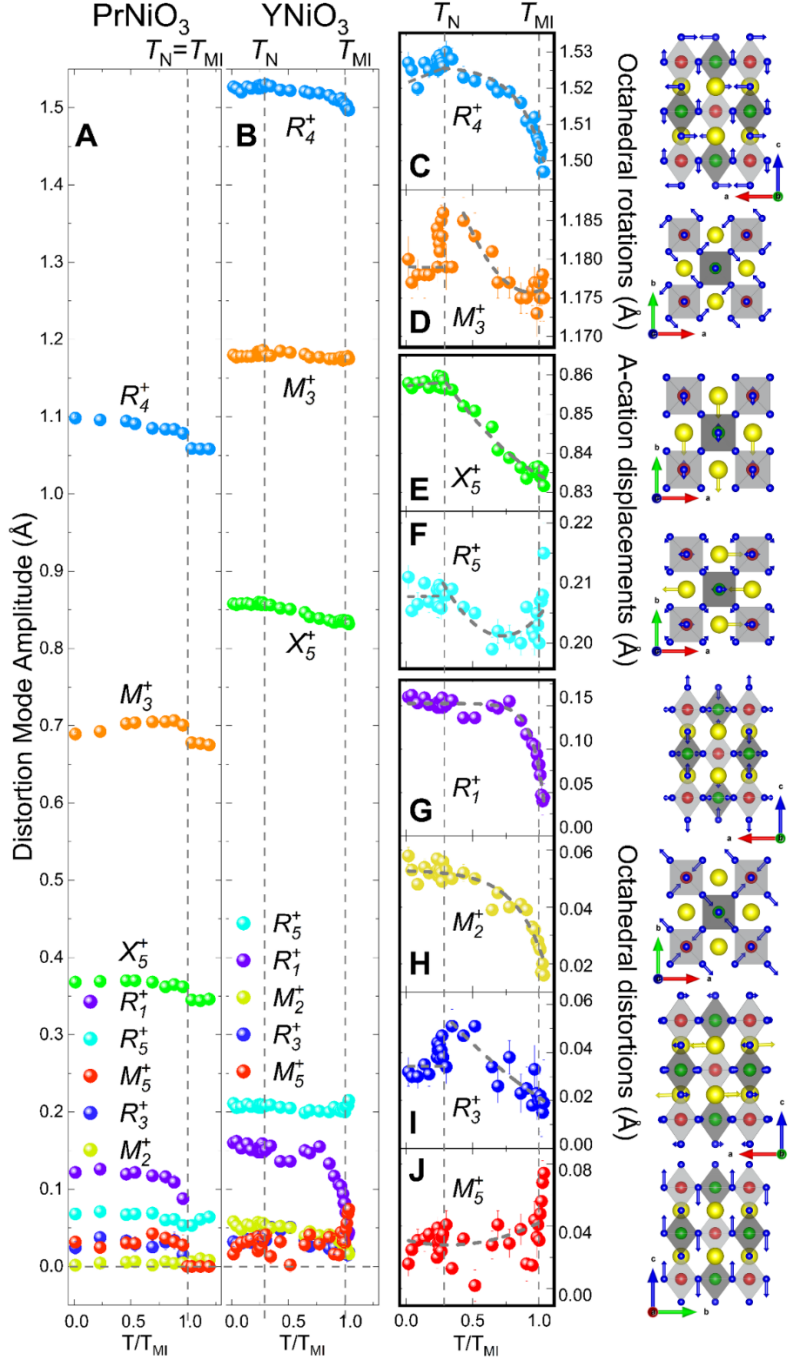

**Figure 4.** Comparison between the mode amplitudes of YNiO$_3$ and PrNiO$_3$. (A, B) Mode amplitudes of PrNiO$_3$ (from ref. [44]) and YNiO$_3$ as a function of the reduced temperature $T/T_{MIT}$ obtained from Rietveld fits of NPD data using the $P2_1/n$ SG. (C-J) Detailed view showing the presence of anomalies in several mode amplitudes at $T_N$ together with schematic representations of the associated atomic displacements.

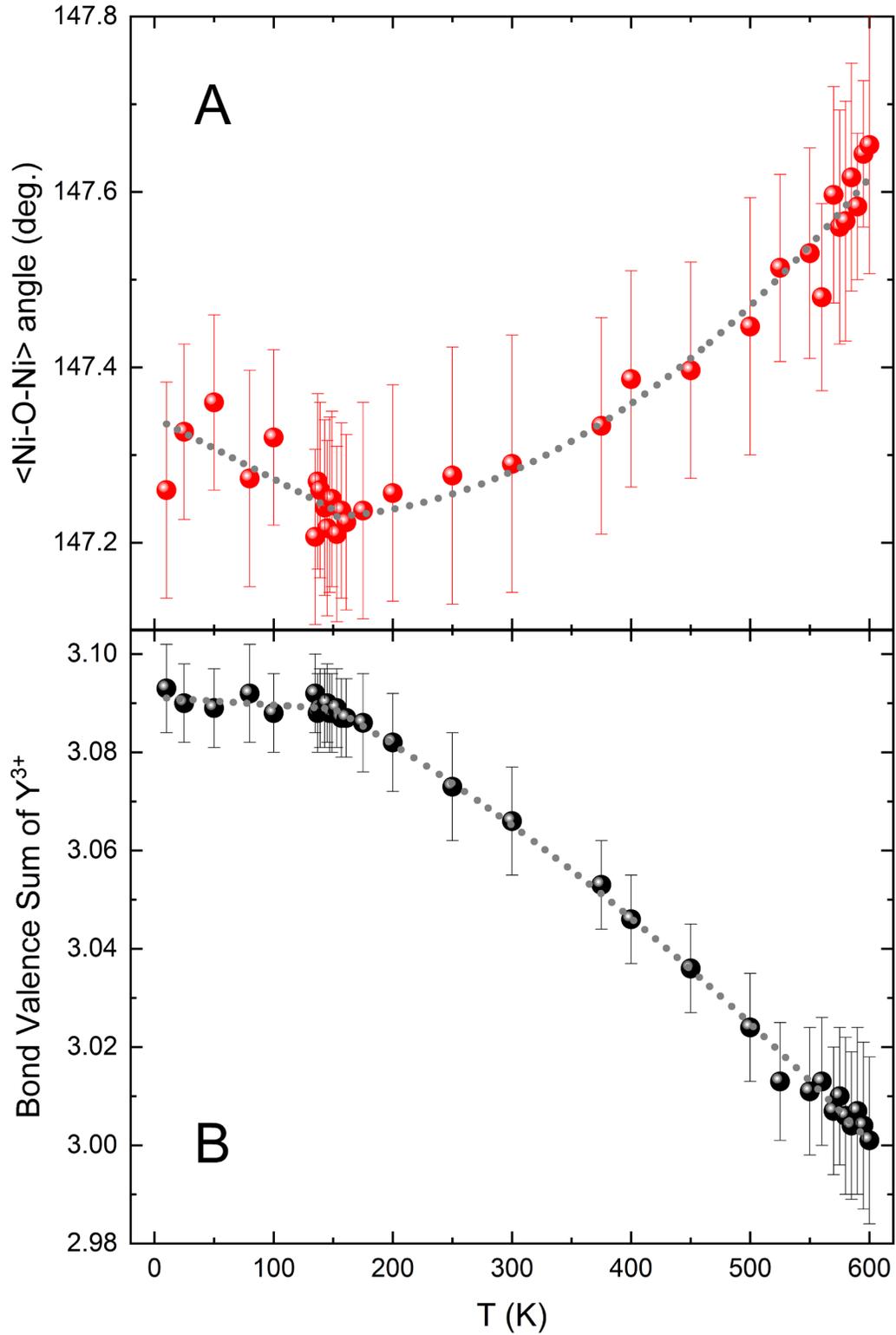

**Figure 5.** Temperature dependence of (A) the average Ni-O-Ni superexchange angle and (B) the $Y^{3+}$ Bond-Valence Sum, showing the existence of anomalies at $T_N$.

Pbnm → 9 modes

| | |
|---|---|
| $R_4^+$ | 1 |
| $M_3^+$ | 1 |
| $X_5^+$ | 2 |
| $R_5^+$ | 4 |
| $M_2^+$ | 1 |

$P2_1/n$ → + 3 modes

| | |
|---|---|
| $R_1^+$ | 1 |
| $M_5^+$ | 1 |
| $M_3^+$ | 1 |

$P2_1$ → + 18 modes

| | |
|---|---|
| $GM_4^-$ | 4 |
| $GM_5^-$ | 2 |
| $R_4^-$ | 1 |
| $X_3^-$ | 2 |
| $X_5^-$ | 3 |
| $M_2^-$ | 1 |
| $M_3^-$ | 2 |
| $M_5^-$ | 3 |

A

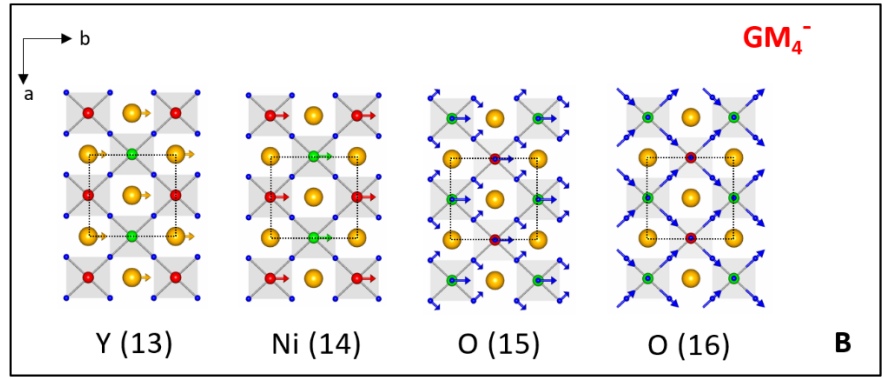

B — $GM_4^-$: Y (13), Ni (14), O (15), O (16)

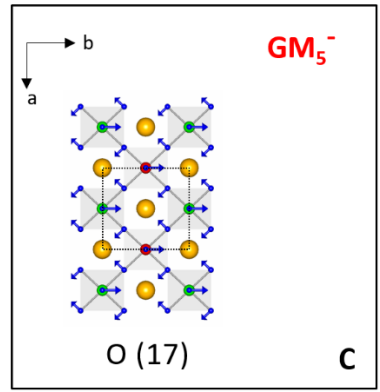

C — $GM_5^-$: O (17)

**Figure 6.** (A) *irrep* labels and number of distortion modes associated with each *irrep* for the space groups *Pbnm*, *P2₁/n* and *P2₁*. (B-C) Schematic representation of the distortion modes of pure *P2₁* symmetry able to give rise to a net polarization component along the **b** crystal axis. (A), Γ-point (irrep *GM4⁻*). (B), R-point (irrep *R4⁻*). For a given mode, the arrows indicate the direction of the associated atomic displacements. The *irrep* and mode labelling is the same as in Table S5.

**AUTHOR INFORMATION**

**Corresponding Authors**

Dariusz Jakub Gawryluk - Laboratory for Multiscale Materials Experiments, Paul Scherrer Institut, CH-5232 Villigen, Switzerland; https://orcid.org/0000-0003-4460-7106;
Email: dariusz.gawryluk@psi.ch.

Yannick Maximilian Klein - Laboratory for Multiscale Materials Experiments, Paul Scherrer Institut, CH-5232 Villigen, Switzerland; Email: maximilian.klein@psi.ch

Marisa Medarde - Laboratory for Multiscale Materials Experiments, Paul Scherrer Institut, CH-5232 Villigen, Switzerland; https://orcid.org/0000-0003-1588-2870
Email: marisa.medarde@psi.ch

**Author Contributions**

The manuscript was written through contributions of all authors. All authors have given approval to the final version of the manuscript. †These authors contributed equally.

**Funding Sources**

Swiss National Science Foundation through the NCCR MARVEL, Grant No. 51NF40-182892 (DJG, YMK, MM); Swiss National Science Foundation R'equip Grant n. 206021_139082 (MM).

**Notes**

The authors declare no competing financial interest.

**ACKNOWLEDGMENTS**

We thank K. Conder, J. Karpinski, P. Lacorre, A. Hampel, C. Ederer, N. Spaldin, A. Georges, O. Peil, D. van der Marel, J. Tessier, I. Ardizzone, Philip Moll, A. Estry, and D. Mazzone for fruitful discussions. We acknowledge the allocation of beam time at the Swiss Spallation Neutron Source SINQ (PSI, Villigen), the Swiss Light Source SLS (PSI, Villigen), and the Europen Synchrotron Radiation Facility ESRF (Grenoble, France).

**REFERENCES**

(1) Wigner, E. On the interaction of electrons in metals. *Phys Rev* **1934**, *46*, 1002-1011.
DOI: 10.1103/PhysRev.46.1002.

(2) Thorne, R. E. Charge-density-wave conductors. *Phys Today* **1996**, *49*, 42-47.
DOI: 10.1063/1.881498.

(3) Mazzone, D. G.; Gerber, S.; Gavilano, J. L.; Sibille, R.; Medarde, M.; Delley, B.; Ramakrishnan, M.; Neugebauer, M.; Regnault, L. P.; Chernyshov, D.; et al. Crystal structure and phonon softening in $Ca_3Ir_4Sn_{13}$. *Phys Rev B* **2015**, *92*, 024101-024106.
DOI: 10.1103/PhysRevB.92.024101.

(4) Nocerino, E.; Sanlorenzo, I.; Papadopoulos, K.; Medarde, M.; Lyu, J.; Klein, Y. M.; Minelli, A.; Hossain, Z.; Thamizhavel, A.; Lefmann, K.; et al. Multiple unconventional charge density wave transitions in $LaPt_2Si_2$ superconductor clarified with high-energy X-ray diffraction. *Commun Mater* **2023**, *4*, 77-88.
DOI: 10.1038/s43246-023-00406-y.

(5) Guguchia, Z.; Gawryluk, D. J.; Shin, S.; Hao, Z.; Mielke Iii, C.; Das, D.; Plokhikh, I.; Liborio, L.; Shenton, J. K.; Hu, Y.; et al. Hidden magnetism uncovered in a charge ordered bilayer kagome material $ScV_6Sn_6$. *Nat Commun* **2023**, *14*, 7796.
DOI: 10.1038/s41467-023-43503-9

(6) Plokhikh, I.; Mielke III, C.; Nakamura, H.; Petricek, V.; Qin, Y.; Sazgari, V.; Küspert, J.; Bialo, I.; Shin, S.; Ivashko, O.; et al. Discovery of charge order above room-temperature in the prototypical kagome superconductor $La(Ru_{1-x}Fe_x)_3Si_2$. *Comm. Phys.* **2024**, *7*, 182-189.
DOI: 10.1038/s42005-024-01673-y.

(7) Verwey, E. J. W. Electronic conduction of magnetite ($Fe_3O_4$) and its transition point at low temperatures. *Nature* **1939**, *144*, 327-328.
DOI: 10.1038/144327b0.

(8) Jirak, Z.; Krupicka, S.; Simsa, Z.; Dlouha, M.; Vratislav, S. Neutron-Diffraction Study of $Pr_{1-x}Ca_xMnO_3$ Perovskites. *J Magn Magn Mater* **1985**, *53*, 153-166.
DOI: 10.1016/0304-8853(85)90144-1.

(9) Li, J. Q.; Matsui, Y.; Park, S. K.; Tokura, Y. Charge ordered states in $La_{1-x}Sr_xFeO_3$. *Phys Rev Lett* **1997**, *79*, 297-300.
DOI: 10.1103/PhysRevLett.79.297.

(10) Castellan, J. P.; Rosenkranz, S.; Osborn, R.; Li, Q.; Gray, K. E.; Luo, X.; Welp, U.; Karapetrov, G.; Ruff, J. P. C.; van Wezel, J. Chiral Phase Transition in Charge Ordered 1T-$TiSe_2$. *Phys Rev Lett* **2013**, *110*.
DOI: 10.1103/PhysRevLett.110.196404.

(11) van den Brink, J.; Khomskii, D. I. Multiferroicity due to charge ordering. *J Phys-Condens Mat* **2008**, *20*, 434217.
DOI: 10.1088/0953-8984/20/43/434217.

(12) Choi, Y. J.; Yi, H. T.; Lee, S.; Huang, Q.; Kiryukhin, V.; Cheong, S. W. Ferroelectricity in an Ising chain magnet. *Phys Rev Lett* **2008**, *100*, 047601-047604.
DOI: 10.1103/PhysRevLett.100.047601.

(13) Wu, H.; Burnus, T.; Hu, Z.; Martin, C.; Maignan, A.; Cezar, J. C.; Tanaka, A.; Brookes, N. B.; Khomskii, D. I.; Tjeng, L. H. Ising Magnetism and Ferroelectricity in $Ca_3CoMnO_6$. *Phys Rev Lett* **2009**, *102*, 026404-260407.
DOI: 10.1103/PhysRevLett.102.026404.

(14) Zhang, Y.; Xiang, H. J.; Whangbo, M. H. Interplay between Jahn-Teller instability, uniaxial magnetism, and ferroelectricity in $Ca_3CoMnO_6$. *Phys Rev B* **2009**, *79*, 054432.
DOI: 10.1103/PhysRevB.79.054432.

(15) Shi, J.; Song, J. D.; Wu, J. C.; Rao, X.; Che, H. L.; Zhao, Z. Y.; Zhou, H. D.; Ma, J.; Zhang, R. R.; Zhang, L.; et al. Ferroelectricity of structural origin in the spin-chain compounds $Ca_3Co_{2-x}Mn_xO_6$. *Phys Rev B* **2017**, *96*, 064112-064121.
DOI: 10.1103/PhysRevB.96.064112.

(16) Spaldin, N. A.; Ramesh, R. Advances in magnetoelectric multiferroics. *Nat Mater* **2019**, *18*, 203-212.
DOI: 10.1038/s41563-018-0275-2.

(17) Fiebig, M.; Lottermoser, T.; Meier, D.; Trassin, M. The evolution of multiferroics. *Nat Rev Mater* **2016**, *1*, 16046.
DOI: 10.1038/natrevmats.2016.46.

(18) Alonso, J. A.; García-Muñoz, J. L.; Fernández-Díaz, M. T.; Aranda, M. A. G.; Martínez-Lope, M. J.; Casais, M. T. Charge Disproportionation in $RNiO_3$ Perovskites: Simultaneous Metal-Insulator and Structural Transition in $YNiO_3$. *Phys Rev Lett* **1999**, *82*, 3871-3874.
DOI: 10.1103/PhysRevLett.82.3871.

(19) Alonso, J. A.; Martínez-Lope, M. J.; Casais, M. T.; García-Muñoz, J. L.; Fernández-Díaz, M. T.; Aranda, M. A. G. High-temperature structural evolution of RNiO$_3$ (R = Ho, Y, Er, Lu) perovskites: Charge disproportionation and electronic localization. *Phys Rev B* **2001**, *64*, 094102.
DOI: 10.1103/PhysRevB.64.094102.

(20) Lacorre, P.; Torrance, J. B.; Pannetier, J.; Nazzal, A. I.; Wang, P. W.; Huang, T. C. Synthesis, Crystal-Structure, and Properties of Metallic PrNiO$_3$ - Comparison with Metallic NdNiO$_3$ and Semiconducting SmNiO$_3$. *J Solid State Chem* **1991**, *91*, 225-237.
DOI: 10.1016/0022-4596(91)90077-U.

(21) Medarde, M. L. Structural, magnetic and electronic properties of RNiO$_3$ perovskites (R equals rare earth). *J Phys-Condens Mat* **1997**, *9*, 1679-1707.
DOI: 10.1088/0953-8984/9/8/003.

(22) García-Muñoz, J. L.; Rodríguez-Carvajal, J.; Lacorre, P. Neutron-Diffraction Study of the Magnetic-Ordering in the Insulating Regime of the Perovskites RNiO$_3$ (R=Pr and Nd). *Phys Rev B* **1994**, *50*, 978-992.
DOI: 10.1103/PhysRevB.50.978.

(23) Fernández-Díaz, M. T.; Alonso, J. A.; Martínez-Lope, M. J.; Casais, M. T.; García-Muñoz, J. L. Magnetic structure of the HoNiO$_3$ perovskite. *Phys Rev B* **2001**, *64*, 144417-144421.
DOI: 10.1103/PhysRevB.64.144417.

(24) Lee, S.; Fernández-Díaz, M. T.; Kimura, H.; Noda, Y.; Adroja, D. T.; Lee, S.; Park, J.; Kiryukhin, V.; Cheong, S. W.; Mostovoy, M.; et al. Negative magnetostrictive magnetoelectric coupling of BiFeO$_3$. *Phys Rev B* **2013**, *88*, 060103-060107.
DOI: 10.1103/PhysRevB.88.060103.

(25) Medarde, M.; Dallera, C.; Grioni, M.; Delley, B.; Vernay, F.; Mesot, J.; Sikora, M.; Alonso, J. A.; Martínez-Lope, M. J. Charge disproportionation in RNiO$_3$ perovskites (R=rare earth) from high-resolution x-ray absorption spectroscopy. *Phys Rev B* **2009**, *80*, 245105-245109.
DOI: 10.1103/PhysRevB.80.245105.

(26) Scagnoli, V.; Staub, U.; Mulders, A. M.; Janousch, M.; Meijer, G. I.; Hammerl, G.; Tonnerre, J. M.; Stojic, N. Role of magnetic and orbital ordering at the metal-insulator transition in NdNiO$_3$. *Phys Rev B* **2006**, *73*, 100409-100412.
DOI: 10.1103/PhysRevB.73.100409.

(27) Zhou, J. S.; Goodenough, J. B. Chemical bonding and electronic structure of RNiO$_3$ (R=rare earth). *Phys Rev B* **2004**, *69*, 153105-153108.
DOI: 10.1103/PhysRevB.69.153105.
(28) Green, R. J.; Haverkort, M. W.; Sawatzky, G. A. Bond disproportionation and dynamical charge fluctuations in the perovskite rare-earth nickelates. *Phys Rev B* **2016**, *94*, 195127-195131.
DOI: 10.1103/PhysRevB.94.195127.

(29) Johnston, S.; Mukherjee, A.; Elfimov, I.; Berciu, M.; Sawatzky, G. A. Charge Disproportionation without Charge Transfer in the Rare-Earth-Element Nickelates as a Possible Mechanism for the Metal-Insulator Transition. *Phys Rev Lett* **2014**, *112*, 106404-106408.
DOI: 10.1103/PhysRevLett.112.106404.

(30) Korosec, L.; Pikulski, M.; Shiroka, T.; Medarde, M.; Luetkens, H.; Alonso, J. A.; Ott, H. R.; Mesot, J. New magnetic phase in the nickelate perovskite TiNiO$_3$. *Phys Rev B* **2017**, *95*, 060411-060415.
DOI: 10.1103/PhysRevB.95.060411.

(31) Perez-Mato, J. M.; Gallego, S. V.; Elcoro, L.; Tasci, E.; Aroyo, M. I. Symmetry conditions for type II multiferroicity in commensurate magnetic structures. *J Phys-Condens Mat* **2016**, *28*, 286001.
DOI: 10.1088/0953-8984/28/28/286001.

(32) Demazeau, G.; Marbeuf, A.; Pouchard, M.; Hagenmuller, P. Series of Oxidized Trivalent Nickel Compounds Derived from Perovskite. *J Solid State Chem* **1971**, *3*, 582.
DOI: 10.1016/0022-4596(71)90105-8.

(33) Alonso, J. A.; Martínez-Lope, M. J.; Casais, M. T.; Aranda, M. A. G.; Fernández-Díaz, M. T. Metal-insulator transitions, structural and microstructural evolution of RNiO$_3$ (R = Sm, Eu, Gd, Dy, Ho, Y) perovskites: Evidence for room-temperature charge disproportionation in monoclinic HoNiO$_3$ and YNiO$_3$. *J Am Chem Soc* **1999**, *121*, 4754-4762.
DOI: 10.1021/ja984015x.

(34) Zhou, J. S.; Goodenough, J. B.; Dabrowski, B.; Klamut, P. W.; Bukowski, Z. Probing the metal-insulator transition in Ni(III)-oxide perovskites. *Phys Rev B* **2000**, *61*, 4401-4404.
DOI: DOI 10.1103/PhysRevB.61.4401.

(35) Kim, S. J.; Demazeau, G.; Presniakov, I. The stabilization of the highest oxidation states of transition metals under oxygen pressures: TlNiO$_3$, a new Ni(III) perovskite-comparison of the electronic properties with those of TNiO$_3$ (T = rare earth and Y). *J Phys-Condens Mat* **2002**, *14*, 10741-10745.
DOI: 10.1088/0953-8984/14/44/369.

(36) Prakash, J.; Blakely, C. K.; Poltavets, V. V. Low temperature high-pressure synthesis of LnNiO$_3$ (Ln = Eu, Gd) in molten salts. *Solid State Sci* **2013**, *17*, 72-75.
DOI: 10.1016/j.solidstatesciences.2012.12.002.

(37) Li, Z. A.; Yan, F. B.; Li, X. Y.; Cui, Y. C.; Wang, V.; Wang, J. O.; Liu, C.; Jiang, Y.; Chen, N. F.; Chen, J. K. Molten-salt synthesis of rare-earth nickelate electronic transition semiconductors at medium high metastability. *Scripta Mater* **2022**, *207*, 114271.
DOI: 10.1016/j.scriptamat.2021.114271.

(38) Klein, Y. M.; Kozłowski, M.; Linden, A.; Lacorre, P.; Medarde, M.; Gawryluk, D. J. RENiO$_3$ single crystals (RE = Nd, Sm, Gd, Dy, Y, Ho, Er, Lu) grown from molten salts under 2000 bar oxygen-gas pressure. *Crystal Growth & Design* **2021**, *21*, 4230.
DOI: 10.1021/acs.cgd.1c00474.

(39) Ardizzone, I.; Teyssier, J.; Crassee, I.; Kuzmenko, A. B.; Mazzone, D. G.; Gawryluk, D. J.; Medarde, M.; van der Marel, D. Raman spectroscopic evidence for multiferroicity in rare earth nickelate single crystals. *Physical Review Research* **2021**, *3*, 033007.
DOI: 10.1103/PhysRevResearch.3.033007.

(40) Alonso, J. A.; Martínez-Lope, M. J.; Casais, M. T.; García-Muñoz, J. L.; Fernández-Díaz, M. T. Room-temperature monoclinic distortion due to charge disproportionation in RNiO$_3$ perovskites with small rare-earth cations (R = Ho, Y, Er, Tm, Yb, and Lu): A neutron diffraction study. *Phys Rev B* **2000**, *61*, 1756-1763.
DOI: 10.1103/PhysRevB.61.1756.

(41) Perez-Mato, J. M.; Orobengoa, D.; Aroyo, M. I. Mode crystallography of distorted structures. *Acta Crystallogr A* **2010**, *66*, 558-590. DOI: 10.1107/S0108767310016247.

(42) Gibbs, A. S.; Knight, K. S.; Lightfoot, P. High-temperature phase transitions of hexagonal YMnO. *Phys Rev B* **2011**, *83*, 094111-094119.
DOI: 10.1103/PhysRevB.83.094111.

(43) Senn, M. S.; Bombardi, A.; Murray, C. A.; Vecchini, C.; Scherillo, A.; Luo, X.; Cheong, S. W. Negative Thermal Expansion in Hybrid Improper Ferroelectric Ruddlesden-Popper Perovskites by Symmetry Trapping. *Phys Rev Lett* **2015**, *114*, 035701-035705.
DOI: 10.1103/PhysRevLett.114.035701.

(44) Gawryluk, D. J.; Klein, Y. M.; Shang, T.; Sheptyakov, D.; Keller, L.; Casati, N.; Lacorre, P.; Fernández-Díaz, M. T.; Rodríguez-Carvajal, J.; Medarde, M. Distortion mode anomalies in bulk PrNiO$_3$: Illustrating the potential of symmetry-adapted distortion mode analysis for the study of phase transitions. *Phys Rev B* **2019**, *100*, 205137.
DOI: 10.1103/PhysRevB.100.205137.

(45) Blasco, J.; Subías, G.; García-Muñoz, J. L.; Fauth, F.; García, J. Determination of the Crystal Structures in the A-Site-Ordered YBaMn$_2$O$_6$ Perovskite. *J Phys Chem C* **2021**, *125*, 19467-19480.
DOI: 10.1021/acs.jpcc.1c04697.

(46) Harrison, W. A. *The Electronic Structure and Properties of Solids Freeman: the physics of the chemical bond*; W.H. Freeman and Co., 1980.
DOI: 10.1016/0022-2860(81)85136-8.

(47) Moskvin, A. S. Dzyaloshinskii Interaction and Exchange-Relativistic Effects in Orthoferrites. *J. Exp. Theor. Phys.* **2021**, *132*, 517-547.
DOI: 10.1134/S1063776121040245

(48) Zhou, J. S.; Goodenough, J. B.; Dabrowski, B. Exchange interaction in the insulating phase of $RNiO_3$. *Phys Rev Lett* **2005**, *95*, 127204-127207.
DOI: 10.1103/PhysRevLett.95.127204.

(49) Park, K.; Sim, H.; Leiner, J. C.; Yoshida, Y.; Jeong, J.; Yano, S.; Gardner, J.; Bourges, P.; Klicpera, M.; Sechovsky, V.; et al. Low-energy spin dynamics of orthoferrites $AFeO_3$ (A = Y, La, Bi). *J Phys-Condens Mat* **2018**, *30*, 235802.
DOI: 10.1088/1361-648X/aac06b.

(50) Jeong, J.; Goremychkin, E. A.; Guidi, T.; Nakajima, K.; Jeon, G. S.; Kim, S. A.; Furukawa, S.; Kim, Y. B.; Lee, S.; Kiryukhin, V.; et al. Spin Wave Measurements over the Full Brillouin Zone of Multiferroic $BiFeO_3$. *Phys Rev Lett* **2012**, *108*.
DOI: 10.1103/PhysRevLett.108.077202.

(51) Argyriou, D. N.; Mitchell, J. F.; Radaelli, P. G.; Bordallo, H. N.; Cox, D. E.; Medarde, M.; Jorgensen, J. D. Lattice effects and magnetic structure in the layered colossal magnetoresistance manganite $La_{2-2x}Sr_{1+2x}Mn_2O_7$, x = 0.3. *Phys Rev B* **1999**, *59*, 8695-8702.
DOI: 10.1103/PhysRevB.59.8695.

(52) Medarde, M.; Mitchell, J. F.; Millburn, J. E.; Short, S.; Jorgensen, J. D. Optimal $T_C$ in layered manganites:: Different roles of coherent and incoherent lattice distortions. *Phys Rev Lett* **1999**, *83*, 1223-1226.
DOI: 10.1103/PhysRevLett.83.1223.

(53) Chapon, L. C.; Blake, G. R.; Gutmann, M. J.; Park, S.; Hur, N.; Radaelli, P. G.; Cheong, S. W. Structural anomalies and multiferroic behavior in magnetically frustrated $TbMnO_3$. *Phys Rev Lett* **2004**, *93*, 177402-177405.
DOI: 10.1103/PhysRevLett.93.177402.

(54) Fabrèges, X.; Petit, S.; Mirebeau, I.; Pailhès, S.; Pinsard, L.; Forget, A.; Fernandez-Diaz, M. T.; Porcher, F. Spin-Lattice Coupling, Frustration, and Magnetic Order in Multiferroic $RMnO_3$. *Phys Rev Lett* **2009**, *103*, 067204-067207.
DOI: 10.1103/PhysRevLett.103.067204.

(55) Azimonte, C.; Granado, E.; Terashita, H.; Park, S.; Cheong, S. W. Polar atomic displacements in multiferroics observed via anomalous x-ray diffraction. *Phys Rev B* **2010**, *81*, 012103-012106.
DOI: 10.1103/PhysRevB.81.012103.

(56) Damay, F.; Sottmann, J.; Fauth, F.; Suard, E.; Maignan, A.; Martin, C. High temperature spin-driven multiferroicity in ludwigite chromocuprate $Cu_2BrBO_5$. *Appl Phys Lett* **2021**, *118*, 192903.
DOI: 10.1063/5.0049174.

(57) Tokunaga, M.; Azuma, M.; Shimakawa, Y. High-Field Study of Strong Magnetoelectric Coupling in Single-Domain Crystals of BiFeO$_3$. *J Phys Soc Jpn* **2010**, *79*, 064713.
DOI: 10.1143/Jpsj.79.064713.

(58) Park, J.; Lee, S. H.; Lee, S.; Gozzo, F.; Kimura, H.; Noda, Y.; Choi, Y. J.; Kiryukhin, V.; Cheong, S. W.; Jo, Y.; et al. Magnetoelectric Feedback among Magnetic Order, Polarization, and Lattice in Multiferroic BiFeO$_3$. *J Phys Soc Jpn* **2011**, *80*, 114714.
DOI: 10.1143/JPSJ.80.114714.

(59) Girardot, C.; Kreisel, J.; Pignard, S.; Caillault, N.; Weiss, F. Raman scattering investigation across the magnetic and metal-insulator transition in rare earth nickelate RNiO$_3$ (R=Sm, Nd) thin films. *Phys Rev B* **2008**, *78*, 104101.
DOI: 10.1103/PhysRevB.78.104101.

(60) Kumar, H.; Singh, A.; Martinez, J. L.; Alonso, J. A.; Tripathi, S. Unexplored signatures of magnetoelastic and isosymmetric metal-insulator phase transition in a rare-earth nickelate via mode crystallography. *Phys Rev B* **2022**, *106*, 214103-214110.
DOI: 10.1103/PhysRevB.106.214103.

(61) Karpinski, J. High pressure in the synthesis and crystal growth of superconductors and III–N semiconductors. *Philosophical Magazine* **2012**, *92*, 2662-2685.
DOI: 10.1080/14786435.2012.670735.

(62) Willmott, P. R.; Meister, D.; Leake, S. J.; Lange, M.; Bergamaschi, A.; Böge, M.; Calvi, M.; Cancellieri, C.; Casati, N.; Cervellino, A.; et al. The Materials Science beamline upgrade at the Swiss Light Source. *Journal of Synchrotron Radiation* **2013**, *20*, 667-682.
DOI: 10.1107/S0909049513018475.

(63) Bergamaschi, A.; Cervellino, A.; Dinapoli, R.; Gozzo, F.; Henrich, B.; Johnson, I.; Kraft, P.; Mozzanica, A.; Schmitt, B.; Shi, X. T. The MYTHEN detector for X-ray powder diffraction experiments at the Swiss Light Source. *Journal of Synchrotron Radiation* **2010**, *17*, 653-668.
DOI: 10.1107/S0909049510026051.

(64) Fischer, P.; Frey, G.; Koch, M.; Könnecke, M.; Pomjakushin, V.; Schefer, J.; Thut, R.; Schlumpf, N.; Bürge, R.; Greuter, U.; et al. High-resolution powder diffractometer HRPT for thermal neutrons at SINQ. *Physica B* **2000**, *276*, 146-147.
DOI: 10.1016/S0921-4526(99)01399-X.

(65) Diffraction, R. O. *CrysAlisPro Software System, Rigaku Corporation, Wroclaw, Poland,* **2020**. There is no coresponding record for this reference.

(66) Dyadkin, V.; Pattison, P.; Dmitriev, V.; Chernyshov, D. A new multipurpose diffractometer PILATUS@SNBL. *Journal of Synchrotron Radiation* **2016**, *23*, 825-829.
DOI: 10.1107/S1600577516002411.

(67) Rodríguez-Carvajal, J. Recent advances in magnetic structure determination by neutron powder diffraction. *Physica B: Condensed Matter* **1993**, *192*, 55-69.
DOI: 10.1016/0921-4526(93)90108-I.

(68) Altermatt, D.; Brown, I. D. The Automatic Searching for Chemical-Bonds in Inorganic Crystal-Structures. *Acta Crystallogr B* **1985**, *41*, 240-244.
DOI: 10.1107/S0108768185002051.

(69) Brown, I. D.; Altermatt, D. Bond-Valence Parameters Obtained from a Systematic Analysis of the Inorganic Crystal-Structure Database. *Acta Crystallogr B* **1985**, *41*, 244-247.
DOI: 10.1107/S0108768185002063.

(70) Various authors, *International Tables of Crystallography, Volume A*; Edited by M.I. Aroyo Wiley, 2016.
DOI: 10.1107/97809553602060000114.